\documentclass[twocolumn,showpacs,pra,superscriptaddress, amsmath, amssymb,sort&compress 10pt]{revtex4-2}\usepackage[utf8]{inputenc}
\usepackage[sort&compress]{natbib}

\usepackage{qcircuit}
\usepackage{multirow}
\usepackage[pdftex,dvipsnames]{xcolor}
\usepackage{xargs} 
\usepackage{hyperref}

\usepackage[colorinlistoftodos,prependcaption,textsize=tiny]{todonotes}
\newcommandx{\unsure}[2][1=]{\todo[linecolor=red,backgroundcolor=red!25,bordercolor=red,#1]{#2}}
\newcommandx{\change}[2][1=]{\todo[linecolor=blue,backgroundcolor=blue!25,bordercolor=blue,#1]{#2}}
\newcommandx{\info}[2][1=]{\todo[linecolor=OliveGreen,backgroundcolor=OliveGreen!25,bordercolor=OliveGreen,#1]{#2}}
\newcommandx{\improvement}[2][1=]{\todo[linecolor=Plum,backgroundcolor=Plum!25,bordercolor=Plum,#1]{#2}}
\newcommandx{\thiswillnotshow}[2][1=]{\todo[disable,#1]{#2}}
\usepackage{graphicx}
\usepackage{ulem}
\usepackage{hyperref}
\usepackage{orcidlink}
\usepackage{braket} 

\begin{document}
\title{Surrogate optimization of variational quantum circuits}
\author{Erik J. Gustafson${}^{\orcidlink{0000-0001-7217-5692}}$}
\affiliation{USRA Research Institute for Advanced Computer Science (RIACS), NASA Ames Research Center}
\author{Juha Tiihonen${}^{\orcidlink{0000-0003-2107-911X}}$}
\affiliation{Department of Physics, Nanoscience Center, University of Jyv\"{a}skyl\"{a}, Jyv\"{a}skyl\"{a}, Finland}
\author{Diana Chamaki}
\affiliation{Department of Chemistry, Columbia University, New York, New York 10027, USA}
\author{Farshud Sorourifar${}^{\orcidlink{0000-0002-1611-8658}}$}
\affiliation{USRA Research Institute for Advanced Computer Science (RIACS), NASA Ames Research Center}
\affiliation{Department of Chemical and Biomolecular Engineering, The Ohio State University}
\author{J. Wayne Mullinax${}^{\orcidlink{0000-0001-9402-9442}}$}
\affiliation{KBR, Inc., Intelligent Systems Division, NASA Ames Research Center, Moffet Field, CA  94035, USA}
\author{Andy C. Y. Li${}^{\orcidlink{0000-0003-4542-3739}}$}
\affiliation{Fermi National Accelerator Laboratory, Batavia, IL,  60510, USA}
\author{Filip B. Maciejewski${}^{\orcidlink{0000-0002-8503-1193}}$}
\affiliation{USRA Research Institute for Advanced Computer Science (RIACS), NASA Ames Research Center}
\affiliation{Quantum Artificial Intelligence Laboratory (QuAIL), NASA Ames Research Center, CA, USA}
\author{Nicolas PD Sawaya${}^{\orcidlink{0000-0001-8510-848}}$}
\affiliation{Azulene Labs, San Francisco, CA 94115; \textit{and} HPI Biosciences, Oakland, CA 94608  }
\author{Jaron T. Krogel}
\affiliation{ Materials Science and Technology Division, Oak Ridge National Laboratory, Oak Ridge, Tennessee 37831, USA}
\author{David E. Bernal Neira${}^{\orcidlink{0000-0002-8308-5016}}$}
\affiliation{USRA Research Institute for Advanced Computer Science (RIACS), NASA Ames Research Center}
\affiliation{Davidson School of Chemical Engineering,
Purdue University}
\author{Norm M. Tubman}
\email{norman.m.tubman@nasa.gov}
\affiliation{Applied Physics Group,
NASA Ames Research Center}
\begin{abstract}
Variational quantum eigensolvers are touted as a near-term algorithm capable of impacting many applications.  However, the potential has not yet been realized, with few claims of quantum advantage and high resource estimates, especially due to the need for optimization in the presence of noise.  Finding algorithms and methods to improve convergence is important to accelerate the capabilities of near-term hardware for VQE or more broad applications of hybrid methods in which optimization is required. 
To this goal, we look to use modern approaches developed in circuit simulations and stochastic classical optimization, which can be combined to form a surrogate optimization approach to quantum circuits.
Using an approximate (classical CPU/GPU) state vector simulator as a surrogate model, we efficiently calculate an approximate Hessian, passed as an input for a quantum processing unit or exact circuit simulator. This method will lend itself well to parallelization across quantum processing units. 
We demonstrate the capabilities of such an approach with and without sampling noise and a proof-of-principle demonstration on a quantum processing unit utilizing 40 qubits.
\end{abstract}
\date{\today}
\maketitle

\section{Introduction}
\label{sec:punchmewiththeawesomeness}

Preparing accurate ground states for quantum systems on quantum hardware is crucial for the fields of high-energy and atomic physics \cite{delgado2022unsupervised, liu2020quantum, humble2022snowmass, chan2021sissa, Bauer:2022hpo, Funcke:2021aps,avkhadiev2020accelerating,culver2021quantum,ciavarella2022preparation,martyn2022variational,PhysRevLett.79.2586,sqms2022,Gustafson_2023,gusta2022} , biology \cite{Cordier2022,Fedorov2021}, medicine \cite{Genomics, malone2022towards,mathur2022medical,izsak2023quantum}, condensed matter \cite{sawaya2023,magann2023randomized,PhysRevA.104.032610,smith2023deterministic,Unmuth_Yockey_2022,Sherbert_2021,PhysRevResearch.5.033071,PhysRevA.106.022434,xu2020test,kokail2019self,vogt2020preparing,gyawali2022adaptive,bravo2020scaling,khan2023preoptimizing,bassman2022}, and quantum chemistry \cite{TILLY20221,sawaya2023,2014NatCo...5.4213P,2016NJPh...18b3023M,anastasiou2022tetrisadaptvqe,burton2022exact,2021APS..MARS34006C,PRXQuantum.2.020310,Chamaki:2022pjp,2019QS&T....4a4008R,Tilly_2022,Cao_2019,Kandala_2017,chamaki2022selfconsistent,Huggins_2020,chivilikhin2020mogvqe,2020arXiv200212902U,2014arXiv1411.4028F,hempel2018quantum,feniou2023overlap,shen2023,shen2023-1,krem2023,klymko2022,krem2021,krem2022, Lanyon2010}. 
The large unknown costs of constructing these states on both near-term and fault-tolerant hardware is an open area of research \cite{doi:10.1021/jz501649m,2014NatSR...4E6603B,2009JChPh.130s4105W, Tubman2018}.
While fault-tolerant hardware is still in development, it is critical that short-depth efficient circuits for state preparation are available. The variational quantum eigensolver (VQE) has shown promise for moderately sized systems to efficiently construct the states where the circuits are optimized on the quantum hardware. 

VQEs are not without their faults; as the number of parameters increases, these algorithms can become costly and can have exponential scaling in the worst situations, such as in cases in which barren plateaus form~\cite{wang2021noise,Anschuetz_2022, McClean_2018, Cerezo2021, ragone2023unified}. 
Other components of the algorithm must be carefully considered, such as numerical gradient evaluations \cite{bittel2021training, Bri2022}, and choice of ansatz \cite{Thanasilp_2023,PRXQuantum.3.010313,arrasmith2022equivalence,PhysRevX.11.041011,Uvarov_2021,PhysRevResearch.3.033090}.
Many ideas exist to improve and accelerate the reach of VQAs as it is an active field of research \cite{bittel2022fast,PhysRevA.98.032309,Harrow2021,jones2020efficient,Berg2022, zhew2020,PhysRevA.107.032415}.
\begin{figure*}
    \includegraphics[width=6.25in]{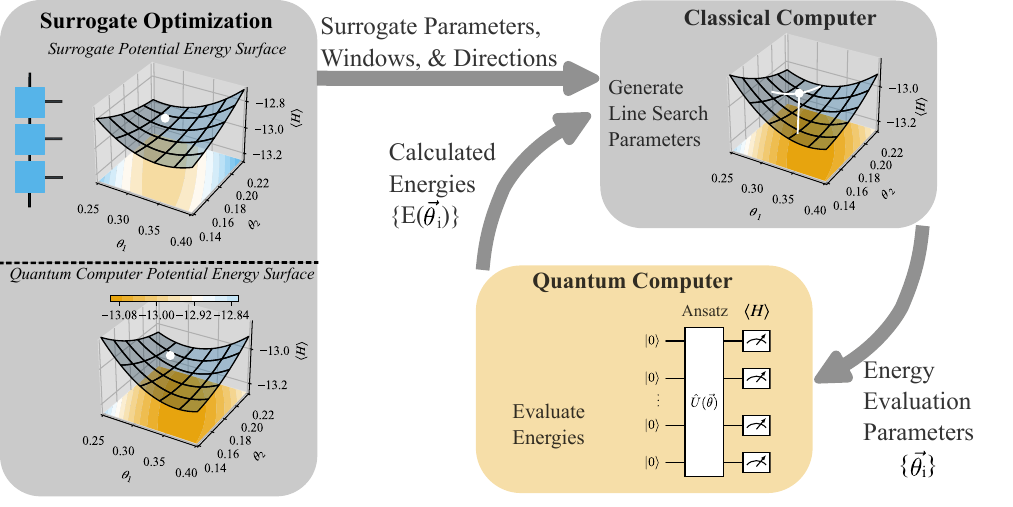}
    \caption{Diagram of the surrogate line search optimization procedure. The energy surfaces shown correspond to a 2D projection of the energy surface for the Ising model studied in this work. The feedback loop between the classical computer and the quantum processing unit (QPU) or quantum computer repeats for as many iterations as desired.}
    \label{fig:daprocedure}
\end{figure*}

An option to bypass these issues is to adopt an approach in which no optimization is performed on the quantum hardware \cite{baek2023say}. One uses heuristic classical computational approaches to optimize the parameters of a quantum circuit before running on actual hardware. Many different approaches can be realized, one of which has been extensively explored with chemistry applications using approximate circuit simulators using up to 64 qubits with moderate computational resources \cite{mullinax2023large, Hirsbrunner:2023wwj}.

The no-optimization approach then leaves the question of how to go beyond the classically optimized circuit effectively to find quantum advantage.   In particular, after optimizing classically as much as possible, what is the best way to go forward and optimize the circuit further with quantum resources? To answer this question, we take inspiration from a geometry optimization algorithm developed for noisy electronic structure calculations \cite{Krogel2022}. 
In \cite{Krogel2022}, the authors use a surrogate model (in this case, density functional theory) to calculate a Hessian for the density functional potential energy surface, which is expected to be near the exact minimum geometry. The approximate Hessian is then used to provide conjugate directions for a line search using a noisy evaluation routine (diffusion Monte Carlo calculations); each line search in a given conjugate direction uses a fixed number of points. In the following work, we translate this to circuit optimization for quantum computing applications. We describe an approach that uses an approximate circuit simulator \cite{mullinax2023large,Hirsbrunner:2023wwj} to obtain an approximate Hessian for the full VQE simulation.

In this work, we show how to apply the surrogate Hessian algorithm to the optimization of quantum circuits. Case studies for a selection of molecules and quantum spin models will demonstrate this approach's effectiveness, including using an IBM quantum computer for the transverse Ising model using 40 qubits.

\section{Theory and Models}
\label{sec:theory}

In this work we employ an optimization procedure laid out in Ref. \cite{Krogel2022}, which optimizes an accurate and noisy cost function with the help of a cheap and noiseless surrogate. The surrogate model is an approximation, but it provides reasonable guesses for the most important features of the optimization landscape, such as a starting point and a Hessian at the minimum. Eigenvectors of the (surrogate) Hessian form a basis of conjugate directions along which the optimization can be solved instantaneously, in principle. Each of the conjugate directions can be optimized in parallel, and thus, a multi-parameter optimization of any size can be treated in a fast-converging series of parallel iterations. Typically, if the surrogate is a good approximation to the high-level function, only a few parallel iterations are necessary. The conjugate directions are solved using a line-search approach, which is robust to noise and independent of gradients: The function parameters are offset by a regular grid of $N$ shifts along the line, and then evaluated using the high-level theory. The minimum is located by fitting a low-order polynomial, and the parameters are shifted accordingly. The surrogate Hessian is kept fixed throughout the iteration. The next iteration uses the same conjugate directions and search window sizes since the Hessian of the low-level theory is fixed. The entire procedure is illustrated in Fig. \ref{fig:daprocedure}.


\begin{table*}
\caption{Surrogate parameters for noisy simulations. $N_{\rm{shots}}(\delta E)$ corresponds to the number of shots per function call required for a desired energy precision. $N_{\rm{PAR}}$ is the number of parameters in the ansatz, $N_{\rm{QUB}}$ is the effective number of qubits, and $N_{\rm{TOT}}$ is the total number of wavefunction determinants.}
\label{tab:surrparsnoise}
\begin{tabular}{cccccccccc}
\hline\hline
Molecule & Basis & $N_{\rm{PAR}}$ & $N_{\rm{QUB}} $ & $N_{\rm{CUT}}$ & $N_{\rm{MAX}}$ & $N_{\rm{TOT}}$& $N_{\rm{shots}}(\delta E = 10^{-3})$ & $N_{\rm{shots}}(\delta E = 10^{-4})$ & $N_{\rm{shots}}(\delta E = 10^{-5})$ \\
\hline
$\rm{H}_2\rm{O}$ & STO-3G & 28 & 14 & 10 & 10 & 1225 & $7\times 10^{5}$ & $7\times 10^{7}$ & $7\times 10^{9}$ \\
$\rm{N}_2$ & STO-3G & 55 & 20 & 60 & 80 &  14,400 & $4\times 10^{6}$ & $4\times 10^{8}$ & $4\times 10^{10}$\\
$\rm{N}_2$ & cc-pVDZ & 50 & 36 & 200-500 & 200-500 & $3\times10^7$ & $2\times10^{8}$ & $2\times 10^{10}$ & $2\times10^{12}$ \\
$\rm{H}_4$ & cc-pVDZ & 193 & 40 & 50 & 50 & 36,100 & $4\times10^{5}$ & $4\times 10^{7}$ & $4\times10^{9}$\\
\hline\hline
\end{tabular}
\end{table*}

%
%

The low-level surrogate, which is used to compute the Hessian in this work, has multiple forms, including a sparse wave function simulator \cite{mullinax2023large, chenucc} and a tensor network simulator \cite{khan2023preoptimizing}. 
The low-level Hessian computed with our sparse wave function simulator (SWS) takes advantage of the sparsity of the electronic wave function to simulate chemical systems of up to 64 qubits that would otherwise be too costly \cite{mullinax2023large}.
The SWS makes large calculations tractable by truncating the wave function, which is controlled by two input parameters, $N_{\rm{MAX}}$ and $N_{\rm{CUT}}$.
After applying each operator in the factorized unitary coupled cluster ansatz \cite{doi:10.1021/acs.jctc.0c01052}, we truncate the wave function to keep only the $N_{\rm{CUT}}$ determinants, or computational basis states, with the largest magnitudes of the wave function amplitudes if the wave function contains more than $N_{\rm{MAX}}$ determinants. This approximate treatment allows us to keep the significant contributions of the wave function and maintain a computationally tractable calculation.  This type of approximation has been used to develop classical algorithms previously~\cite{Tubman:2020:2139,tubman2016,Tubman2018,tubman2018-1,levine2020,dbwy2023}.
In addition, we allow for sampling noise to be included in the simulations.

For this work, we look at two classes of models: molecular Hamiltonians and the transverse field Ising model. The 
molecular Hamiltonians
are written in the form:
\begin{equation}
    \label{eq:bhham}
    \hat{H} = \sum_{i, a}^{\rm{orb.}}c_i^a (\hat{a}^{\dagger}_a \hat{a}_i + h.c.) + \sum_{i,j, a, b}^{\rm{orb.}}c_{ij}^{ab}( \hat{a}^{\dagger}_a \hat{a}^{\dagger}_b \hat{a}_j \hat{a}_i + h.c), 
\end{equation}
where the coefficients $c_i^a,$ and $c_{ij}^{ab}$ are a function of the constituent atoms and their positions. The primary molecules investigated were H$_2$O and N$_2$ in the STO-3g basis, and H$_4$ chain and N$_2$ in the cc-pvdz basis. 
The $\rm{H}_4$ simulations use an intermolecular distance of 1.27$\AA$ while all other electronic structures are taken from the NIST computational chemistry database \cite{NIST_chem_data}.
The H$_4$ molecule uses a stretched interatomic distance because it is a more strongly correlated distance. 
We provide information regarding the surrogate parameters in Tab. \ref{tab:surrparsnoise}. 

The transverse field Ising model we investigated in this work has a Hamiltonian of the form:
\begin{equation}
\label{eq:qpuham}
\hat{H} = J_1 \sum_{i=1}^{N_s}\hat{\sigma}^z_i\hat{\sigma}^z_{i + 1} + J_2 \sum_{i=1}^{N_s}\hat{\sigma}^z_i \hat{\sigma}^z_{i + 2} + h_t \sum_{i=1}^{N_z} \hat{\sigma}^x_i.
\end{equation}
For this model, we set the parameters $J_1 = 1.0$, $J_2 = 0.9$, $h_t = 0.4$, and the number of sites, $N_s=40$,  with periodic boundary conditions. As shown in later sections, benchmark simulations were performed for other system sizes. 

Most of our simulations use the unitary coupled cluster singles and doubles (UCCSD) ansatz, 
\begin{equation}
    \label{eq:uccsd}
    |\Psi_{UCC}\rangle = e^{\hat{T} - \hat{T}^{\dagger}} |\Psi_0\rangle,
\end{equation}
where $|\Psi_0\rangle$ is a reference state, $|\Psi_{UCC}\rangle$ is the target state for the VQE optimization, and $\hat{T}$ is the singles and doubles cluster operator:
\begin{equation}
\label{eq:clusterop}
    \hat{T} = \sum_{i}^{\rm{occ.}}\sum_{a}^{\rm{vir.}}\theta_{i}^{a}\hat{a}^{\dagger}_a\hat{a}_i + \sum_{i,j}^{\rm{occ.}}\sum_{a,b}^{\rm{vir.}}\theta_{ij}^{ab} \hat{a}^{\dagger}_a \hat{a}^{\dagger}_b \hat{a}_j \hat{a}_i .
\end{equation}
The summations in the previous equation extend over the occupied and virtual orbitals. The coefficients, $\theta$, are variational parameters that are optimized so that the energy $E = \langle \Psi_{UCC}|\hat{H}|\Psi_{UCC}\rangle$ is minimized. We use the trotterized form of Eq. (\ref{eq:uccsd}) with the order of the operators according to the coupled cluster singles and doubles amplitudes.

\section{Sampling Noise Studies in Electronic Structure}
\label{sec:results}

\begin{figure*}
    \includegraphics[width=\linewidth]{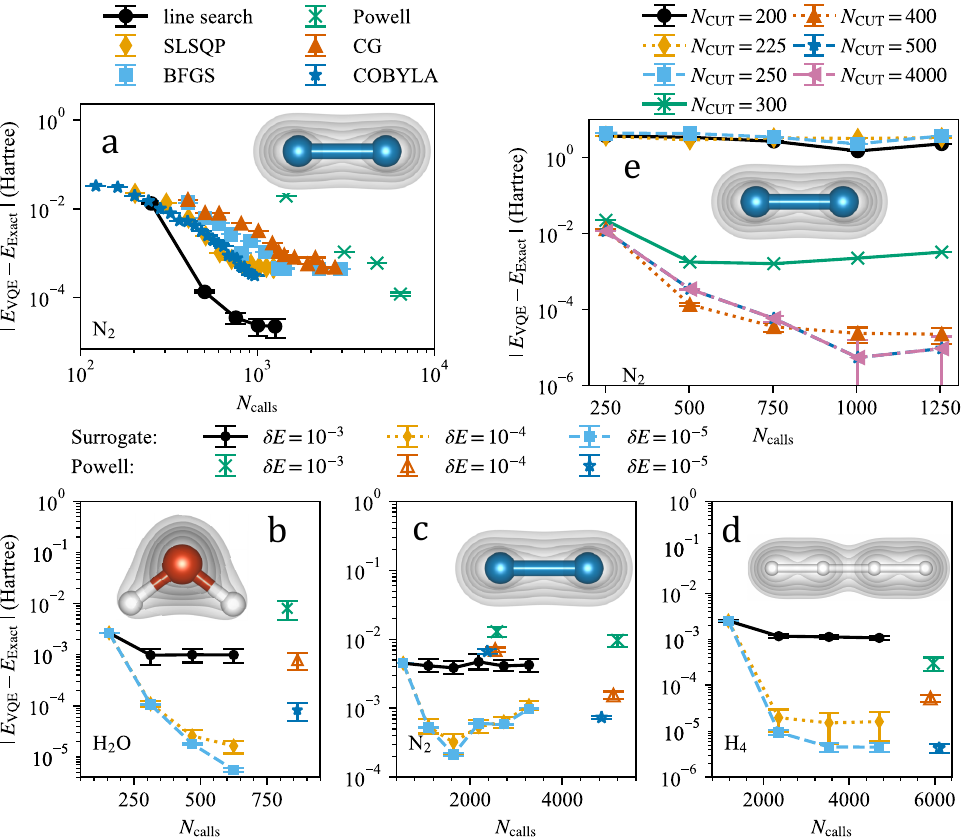}
    \caption{\textbf{a}: A comparison of the performance of surrogate line search, SLSQP, BFGS, Powell, COBYLA, and conjugate gradient (CG) for the N$_2$ molecule in the cc-pvdz basis using the 18 lowest orbitals and 50 terms from the CCSD expansion with the largest coefficients. The resolution on the energy is $\delta E = 10^{-5}$. \textbf{b}-\textbf{d}: Comparison of Powell optimizer to the surrogate line search using 7 points per search direction for three molecules and bases: H$_2$O in STO-3G basis (left), N$_2$ in STO-3G basis (middle), H$_4$ chain in cc-pvdz basis with interatomic distance 1.27$\AA$ (right). Solid markers correspond to the surrogate line search, and open markers correspond to the Powell optimizer. The details of each simulation are provided in Table \ref{tab:surrparsnoise}. \textbf{e}: Comparison of different values of $N_{\rm{CUT}}$ for a truncated $\rm{N}_2$ simulation using the parameters from Table \ref{tab:surrparsnoise} and energy uncertainties  $\delta E = 10^{-5}$.}
    \label{fig:chemicalcomparisonstopowell}
\end{figure*}
In this section, we present results on the feasibility and efficacy of the surrogate line search method. We use the molecular Hamiltonians from Eq. \eqref{eq:bhham} to investigate the feasibility of the surrogate line search method and its performance compared to other traditional optimizers. Furthermore, the effects of surrogate accuracy using the molecule N$_2$ and modifications of the line search algorithm using the H$_2$O molecule are examined. 

In Fig. \ref{fig:chemicalcomparisonstopowell}a, we show a comparison for noisy optimizations of N$_2$ in the cc-pvdz basis using the line search and five traditional classical optimizers: SLSQP, BFGS, Powell, conjugate gradient (CG), and COBYLA. All optimization alternatives start with the same initial parameters determined by the surrogate. 
Compared to the other choices of optimizers, the surrogate line search converges more quickly and with significantly greater accuracy.
These results show that the surrogate line search offers a significant reduction in function calls compared to other classical optimizers.

In light of these results, we investigated the efficacy of the surrogate line search compared to the noise resilient optimizer (Powell) \cite{Singh2023,pellow2021comparison,lockwood2022empirical,powell1964efficient} for three molecules, H$_2$O and N$_2$ in the STO-3G basis, and H$_4$ in the cc-pvdz basis. These simulations were performed using various levels of sampling noise, as discussed below. Additional simulations conducted without noise are provided in the supplemental material.

The Powell and line search methods use initial parameters determined by the surrogate method. The initial search directions for Powell are conjugate directions determined from the approximate optimization's Hessian.
The choices for $N_{\rm{CUT}}$, $N_{\rm{MAX}}$ as well as the number of parameters $N_{\rm{PAR}}$, number of qubits $N_{\rm{QUB}}$, and total number of wavefunction determinants without truncations $N_{\rm{TOT}}$ are provided in Table \ref{tab:surrparsnoise}. 
For each molecule, we consider optimization with error bars (for the energy cost function) $\delta E = \mathcal{O}(10^{-3}),~\mathcal{O}(10^{-4}),\rm{~and~} \mathcal{O}(10^{-5})$. 
The surrogate line search method is then compared with the Powell optimizer. 

We show the comparison between the surrogate line search algorithm and the Powell optimizer in Fig. \ref{fig:chemicalcomparisonstopowell}b-d for $\rm{H}_2\rm{O}$, $\rm{N}_2$, and $\rm{H}_4$. 
For $\rm{H}_2\rm{O}$, we find that the line search converges within three iterations for $\delta E=10^{-3}$ but at least four iterations for higher sampling rates. At least 607 function calls using the surrogate line search algorithm are required to begin converging. In contrast, the Powell optimizer takes between 830 and 850 function calls. Unsurprisingly, as the sampling rate increases, both the Powell and surrogate line search accuracies improve. The Powell optimizations converge after two iterations and agree with the line search. Due to the greater number of function calls, later portions of the Powell optimizations are not shown.   

The $\rm{N}_2$ molecule has almost double the number of parameters as the $\rm{H}_2\rm{O}$ molecule and shows a starker contrast between the two optimization methods. Both optimizers will require more function calls because the number of parameters has increased. As seen in the figures, both methods converge within 2 to 3 iterations; however, the Powell optimizer takes approximately 5000-8000 function calls to converge, whereas the line search only takes 1500-2000, which is a 2.5 to 4 times reduction in cost. The energy can increase because all the search directions are updated simultaneously, as seen in this example. Improved techniques might involve updating search directions sequentially rather than simultaneously. Testing such an approach will be studied in future work, which must be considered regarding the implications for parallelizing the line searches.  

The stretched $\rm{H}_4$ simulation results are also quite stark and are shown in Fig. \ref{fig:chemicalcomparisonstopowell}d.
We expect that the stretched H$_4$ has difficulty with convergence, as it is likely a more strongly correlated system. Convergence to the desired result also happens within 2 to 3 iterations, requiring between 2000 and 4000 function calls. In contrast, the Powell optimizer takes nearly 6000-9000 function calls; this is a cost savings of approximately three times by using the surrogate line search approach.


\begin{figure}[t]
    \centering
    \includegraphics[width=\linewidth]{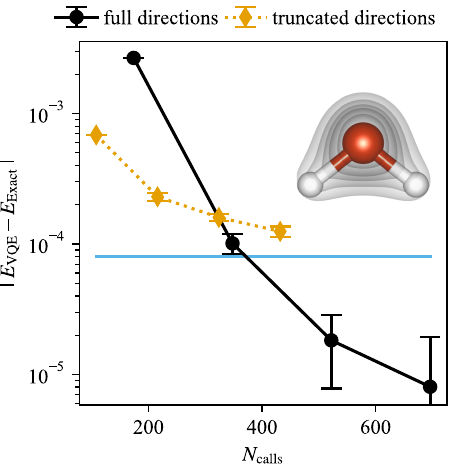}
    \caption{Comparison between the surrogate line search using the 10 steepest of the 29 total set of search directions for H$_2$O in the STO-3G basis versus all available search directions with energy precisions $\delta E = 10^{-4}$. Points displayed are the infinite sampling limit and error bars are calculated using statistical bootstrap from the uncertainties on the parameters. The gold line indicates the optimal result for the truncated search directions.}
    \label{fig:H2otruncs}
\end{figure}

Until now, tunable parameters for the surrogate line search have been fixed for specific problems. An important choice as problem sizes are scaled up is how the accuracy of the surrogate will impact the efficacy of the surrogate line search. 
It will be expected that if the surrogate is too coarse, then changes to the potential energy surface will be quite discontinuous, and estimations of the Hessian, and by analogy, the conjugate search directions, will be negatively affected; in principle, the numerically calculated Hessian could have negative eigenvalues falsely indicating that the system is not in a local minimum.
We show the effects of a line search using values of $N_{\rm{CUT}} = 200,~225,~250,~300,~400,\text{~and}~500$ using the cc-pvdz basis for the N$_2$ in Fig. \ref{fig:chemicalcomparisonstopowell}e. To keep resource costs manageable for this molecule, we use only the 18 lowest orbitals and keep only 50 operators for the UCC ansatz corresponding to the terms from the coupled cluster singles and doubles expansion with the largest coefficients.
For suitably large values of $N_{\rm{CUT}} > 300$, we find that the Hessian is accurate enough to provide a suitable surrogate line search and generally converges to a within $2\times 10^{-3}$ Hartree. However, there is a break-over point, $N_{\rm{CUT}} < 300$, where the Hessian is ill-defined and does not provide accurate search directions. The inaccurate search directions are a by-product of the Hessian being discontinuous in parameter space for the surrogate model. 

Another question we can ask is whether we can reduce the required function calls by limiting the line searches to a subset of the steepest search directions. We look specifically at the STO-3G basis $\rm{H}_2{O}$ simulation for this case. The Hessian calculated using SWS has 10 eigenvalues around 80 and 19 eigenvalues around 5; a larger eigenvalue indicates a steeper search direction. We limit the surrogate line search to the conjugate directions corresponding to the 10 largest eigenvalues. A comparison of this truncated search with the entire search in Fig. \ref{fig:H2otruncs} using a fixed energy error $\delta E=10^{-4}$. The truncated search method converges with fewer function calls than the full line-search. Because the truncated search direction optimization has a lower energy than the full search at early times, this suggests that one could dynamically include search directions as the optimization improves. 

\section{Transverse Ising model using a quantum computer}
\begin{figure}[!b]
\includegraphics[width=\linewidth]{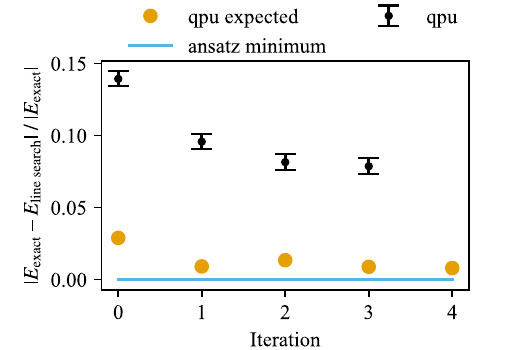}
\caption{40 qubit simulation on \texttt{ibm\_brisbane} using the ansatz in Eq. \eqref{eq:ansatz} using only 2 steepest search directions. The QPU expected points are the results from an MPS simulation with bond dimension 400 using the parameters determined from the corresponding QPU simulation.}
\label{fig:ibmbrisbaneresults}
\end{figure}
\begin{figure}\includegraphics[width=\linewidth]{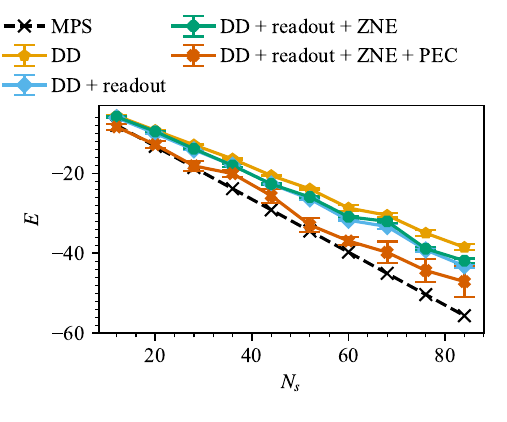}
\vspace{-4em}
    \caption{Energy calculated on $\texttt{ibm\_brisbane}$ using dynamic decoupling (DD),  readout error mitigation (readout), zero noise extrapolation (ZNE), and probabilistic error cancellation (PEC), compared to MPS simulator with bond dimension 40 with parameters fixed from the potential minimum of the surrogate model.}
    \label{fig:noisefloor}
\end{figure}

We used the transverse Ising model from Eq. \eqref{eq:qpuham} as a prototypical example for simulations on quantum processing units (QPUs).
We used this Hamiltonian as an example case to avoid issues measuring long correlated strings of Pauli matrices, such as those appearing in the fermion-to-qubit mapping for quantum chemistry problems.
We use a nearest-neighbor ansatz, which is easily implemented on many quantum platforms. This ansatz uses repeated layers of entangling rotations given by the operator
\begin{equation}
\label{eq:entangling}
\hat{U}_{i,j}(\theta) = e^{-i \theta (\hat{\sigma}^y_i\hat{\sigma}^z_j + \hat{\sigma}^z_i\hat{\sigma}^y_j)}.
\end{equation}
The explicit form of the ansatz used in our work is
\begin{equation}
\label{eq:ansatz}
\begin{split}
    \hat{V}(\vec{\theta}) = & \Bigg(\prod_{i=0}^{N_s / 2}\hat{U}_{2i, 2i + 1}(\theta_1)\Bigg)\Bigg(\prod_{i=0}^{N_s / 2} \hat{U}_{2i + 1, 2i + 2}(\theta_2)\Bigg)\\
    & \Bigg(\prod_{i=0}^{N_s / 2}U_{2i, 2i + 1}(\theta_3)\Bigg)\Bigg(\prod_{i=0}^{N_s / 2} \hat{U}_{2i + 1, 2i + 2}(\theta_4)\Bigg)
    \end{split}
\end{equation}
We performed a surrogate line search on IBM's \texttt{ibm\_brisbane}. The system we simulate uses the parameters $J_1 = 1.0$, $J_2=0.9$, $h_t = 0.4$, and $N_s = 40$ and is in the gapped phase but is close to a phase transition. 

The surrogate model was a matrix product state (MPS) simulator whose maximum bond dimension was set to 4.
After calculating the Hessian using the MPS simulator, we found that two search directions are relatively shallow with eigenvalues of $3\times10^{-2}$ and $4\times10^{-2}$. 
For this reason, we chose to perform the line search across the two steepest search directions. 
The results from the QPU after three iterations are shown in Fig. \ref{fig:ibmbrisbaneresults}. Although the energies extracted from the simulation are well above the expected values, the parameters have converged to 2.5 standard deviations of the expected result and, in the absence of noise, give results within $1\%$ of the ansatz minimum.

Several error mitigation strategies have been leveraged for these simulations, including dynamical decoupling (DD) \cite{PhysRevLett.82.2417, Ezzell2022,Carr1954,Meiboom1958,Maudsley1986,Viola2003}, randomized compilation (RC) \cite{PhysRevA.94.052325,Erhard_2019,li2017efficient, 2018efficienttwirling, 2013PhRvA..88a2314G, 2016efficienttwirling,Silva-PT2008,Winick:2022scr}, and Clifford rescaling \cite{Urbanek2021,Vovrosh:2021ocf,Arahman:2022tkr}. A brief overview of these methods can be found in the supplemental information.
These simulations show that decoherence and amplitude damping are substantial noise sources and that Clifford renormalization does not account for all errors. Each simulation used 20k shots for each basis measurement in the Hamiltonian.

To have confidence in the validity of any simulations, we compared the energy calculated using a variational circuit on a QPU to an MPS representation of the corresponding state for lattices with 12, 20, 24, 28, and 32 sites. We calculate the energy of the $\texttt{ibm\_brisbane}$ QPU using the parameters corresponding to the surrogate minimum. The discrepancy between the surrogate result and various error mitigation techniques is shown in Fig. \ref{fig:noisefloor}, including probabilistic error cancelation  (PEC) \cite{PhysRevLett.119.180509, PhysRevX.8.031027} and zero noise extrapolation (ZNE) \cite{PhysRevLett.119.180509,Kandala_2019, PhysRevX.7.021050}, which were not included in the production calculations. These results are discussed further in the supplemental information. 
There are linear trends in the divergence of the QPU simulations from the expected results. Since the circuit depth does not increase with increased system size, this indicates that the primary source of systematic error is either coherent or depolarizing errors from the entangling gates as opposed to decoherence effects from the qubit lifetime.

\section{Outlook}
\label{sec:outlook}
This work shows that the surrogate Hessian line search method from \cite{Krogel2022} is amenable to the optimization of variational quantum eigensolver circuits in chemistry and condensed matter. 
The line search method has been shown to work both for hardware efficient and unitary coupled cluster type ansatz\"e. 
In particular, this new optimizer outperforms the Powell optimizer by a factor of 2 to 4 in the cases we studied in the presence of sampling noise.
The efficacy of this line search optimization is contingent upon accurate conjugate search directions being obtained from Hessian in the low-level theory;
in particular, if negative eigenvalues appear in the surrogate's Hessian due to discontinuities in parameter space, the line search algorithm struggles to perform an adequate optimization. 
We also investigated the effects of truncating the search directions using only the steepest conjugate directions. We found that the line search can optimize the circuit faster than using all the line search directions at the expense of some degree of accuracy. 
In addition, a proof of principle demonstration was performed on $\texttt{ibm\_brisbane}$ for a transverse Ising model using 40 qubits. 

While the studies here investigate standard example cases, this method opens up the possibility of investigating frustrated quantum spin systems, which are challenging for both tensor network simulators and quantum Monte Carlo methods. Example models encompass a broad range, for example, including Kitaev spin liquids \cite{PhysRevResearch.5.033071, PhysRevA.106.022434, bespalova2021quantum,suzuki2021proximate,banerjee2017neutron}, geometrically frustrated Kagome antiferromagnets \cite{RevModPhys.88.041002,zheng2023unconventional,PhysRevB.106.214429}, and downfolded models of molecules \cite{nakamura2021respack}. 

It is important to investigate methods to mitigate the problems that arise from ill-defined Hessian directions; is it possible to truncate search directions whose Hessian eigenvalues are negative or near zero as a first-pass optimization and then use a traditional optimizer or a modified version of the line search to improve the optimization.

\section{Acknowledgements}
W.M. and N.T. acknowledge funding from the NASA ARMD Transformational Tools and Technology (TTT) Project.
This material is based upon work supported by the U.S. Department of Energy, Office of Science, National Quantum Information Science Research Centers, Superconducting Quantum Materials and Systems Center (SQMS) under contract No. DE-AC02-07CH11359 (N.T.). 
J.T.K. (problem design, surrogate optimization) was supported by the U.S. Department of Energy, Office of Science, Basic Energy Sciences, Materials Sciences and Engineering Division, as part of the Computational Materials Sciences Program and Center for Predictive Simulation of Functional Materials.
E.G., F.B.M., D.B.N, were supported by the NASA Academic Mission Services, Contract No. NNA16BD14C. D.C. and F.S. participated in the Feynman
Quantum Academy internship program.
D.B.N. acknowledges the support of the startup grant of the Davidson School of Chemical Engineering at Purdue University.
The authors would like to thank Sohaib Alam, Stuart Hadfield, Amanda Kahl, Tom Iadecola, Hank Lamm, Aaron Lott, Ruth Van de Water, and Michael Wagman for their helpful comments.
This research used resources of the Oak Ridge Leadership Computing Facility, which is a DOE Office of Science User Facility supported under Contract DE-AC05-00OR22725.
This research used resources of the National Energy Research
Scientific Computing Center, a DOE Office of Science User Facility supported by the Office of Science of the U.S. Department of Energy under Contract No. DE-AC02-05CH11231 using NERSC award ASCR-ERCAP0024469.
We acknowledge the use of IBM Quantum services for this work. The views expressed are those of the authors, and do not reflect the official policy or position of IBM or the IBM Quantum team.

\newpage
\clearpage
\appendix
\onecolumngrid

\section{Sparse Wave Function Simulator}

The sparse wave function simulator (SWS) used for the electronic structure wave function optimizations performed in this work leverages the sparsity of electronic wave functions to simulate systems using up to 64 qubits \cite{mullinax2023large} and is inspired by the adaptive configuration interaction (CI) algorithms developed for classical hardware\cite{Tubman:2020:2139}. Using this approach, we can find good approximations of the electronic energy, gradients, and Hessians using a UCCSD ansatz in a VQE calculation for molecular systems with double zeta and larger basis sets. This is beyond the reach of standard VQE approaches, which typically require minimal basis sets. The SWS is not only appropriate as a low-cost surrogate level of theory described in this manuscript, but it is also a convenient way to benchmark VQE calculations with large basis sets on classical hardware before the availability of suitable quantum hardware as well as a suitable method for generating initial quantum states that can be further refined on quantum hardware.

We employ the factorized form of the UCCSD ansatz given by
\begin{equation}
    \label{eq:ansatz:si}
  \ket{\Psi_{\mathrm{UCCSD}}} = \prod_{I} \hat{U}_{I}\ket{\Psi_{0}}  
\end{equation}
where the UCCSD factors $U_{I}$ are either based on either the single ($\hat{a}_i^a$) or double ($\hat{a}_{ij}^{ab}$) excitation operators of second quantization where orbitals $i$ and $j$ refer to occupied orbitals in the reference wave function $\ket{\Psi_0}$, and orbitals $a$ and $b$ refer to unoccupied orbitals.
The UCCSD single excitation factors are given by
\begin{equation}
    \hat{U}_i^a = \exp{\left[\theta_i^a\left(\hat{a}_i^a - \hat{a}_a^i\right)\right]}
\end{equation}
and the UCCSD double excitation factors are given by
\begin{equation}
    \hat{U}_{ij}^{ab} = \exp{\left[ \theta_{ij}^{ab}\left(\hat{a}_{ij}^{ab} - \hat{a}_{ab}^{ij}\right)\right]}
\end{equation}
Each UCCSD factor can be evaluated using expressions reported by Chen et al.\cite{chenucc} that are assessed efficiently on a classical computer.

Two key approximations account for the efficiency of the SWS that limits (1) the size of the wave function and (2) the size of the ansatz. First, we limit the number of Slater determinants saved during the evaluation of the ansatz. After evaluating each UCCSD factor, we sort the wave function amplitudes by their magnitude. If the number of nonzero amplitudes is greater than the input parameter $N_{\rm{MAX}}$, then we set the amplitudes with the smallest magnitude to zero so that there is only $N_{\rm{CUT}}$ nonzero amplitudes. In this way, we limit the number of nonzero amplitudes we need to keep in memory and the number of determinants we must consider when evaluating each UCCSD factor. Second, we limit the number of UCCSD factors we permit in the ansatz based on the MP2 or CCSD $T_1$ and $T_2$ amplitudes. If we employ an MP2 guess as the initial values of the variational parameters $\theta$, then we order the UCCSD double excitation factors in decreasing order based on the magnitude of the MP2 $T_2$ amplitudes followed by the UCCSD single excitation factors in order based on the orbital indices with the associated parameters set to zero. We limit the size of the UCCSD ansatz by limiting the number of UCCSD double excitation factors to $N_{\rm{Doubles}}$ with the largest magnitude of the initial MP2 amplitudes. When we employ the CCSD initial guess for the variational parameters, we sort the UCCSD factors, both single and double excitations, based on the CCSD $T_1$ and $T_2$ amplitudes instead of strictly separating the doubles and singles as in the initial MP2 guess. To limit the size of the ansatz with the CCSD initial guess, we only keep $N_{\rm{Gates}}$ of the UCCSD factors with the largest magnitude of the initial CCSD amplitudes. We also used point-group symmetry to eliminate the variational parameters that are necessarily zero, which reduces the size of the ansatz that must be considered in the VQE optimization.\cite{Cao:2022:062452}

A new feature added to the SWS from the original simulator described in \cite{mullinax2023large} is the inclusion of sampling errors. One can include Gaussian random noise shifts to calculated energies; this mimics the effects of sampling noise from a quantum computer. A more robust method would involve performing a fermion-to-qubit basis transformation; we found this to be computationally more expensive without yielding a noticeable effect on results.

\section{Surrogate Line Search Algorithm}

The surrogate line search algorithm originally proposed in Ref \cite{Krogel2022} takes a computationally cheap theory, known as the surrogate, and uses it to provide target input information for the more computationally intensive high-level theory.
The surrogate provides \textit{a priori} information that helps speed up and constrain the optimization of the high-level theory. This prior information includes initial conditions, optimized search windows, improved search directions, and efficient sampling weights. All of these priors together help stabilize the optimizer against many noise sources.

Before optimizations can happen using the high-level theory, one needs the initial search directions and optimal search windows. These directions are found by calculating the Hessian of the potential energy surface along all parameter directions; the eigenvalues of the Hessian are measures of the floppiness of the direction, and the eigenvectors are the improved (conjugate) search directions. 
If one finds eigenvalues relatively close to zero, one can, in principle, remove these search directions since they could negatively impact the eventual optimization.

The optimization of the search windows is a slightly trickier task. 
The eigenvalues are used as an initial guess for the curvature of the potential energy surface and relative search window sizes for the line searches. The preoptimization procedure uses some target input precision for either the parameters or the overall energy and attempts to find the search window size for a given tolerance that maximizes the sampling noise. These optimizations are performed over all of the conjugate search directions. 
The search window sizes are then fixed for the remainder of the optimization. 

Finally, one can continue to optimize using the high-level theory. 
A proposed set of $1 + (M - 1) N_{dir}$ sets of parameters are proposed corresponding to $M$ points along each of the $N_{dir}$ search directions. Once determinations of the energies of each point are performed, a low-order polynomial, between cubic and quartic, is fit to the data points along a given search direction.
The minimum of the fit along each direction is stored and used as the starting point for the next round of line searches. 
The line searches are then performed iteratively until a satisfactory convergence is seen.

\begin{figure}
    \includegraphics[width=\linewidth]{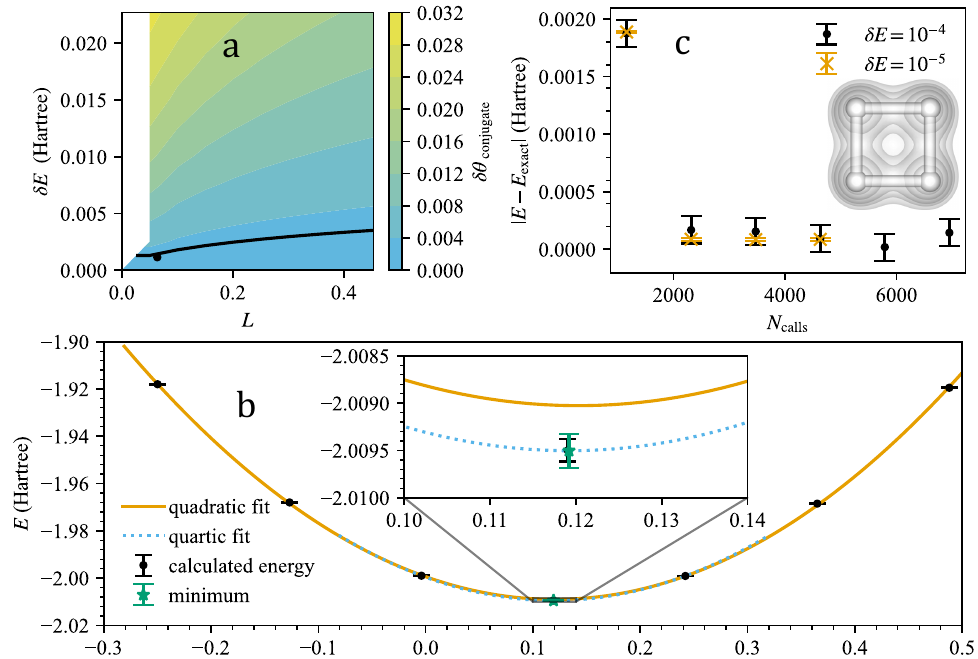}
    \caption{\textbf{a}:  Isocurves of constant parameter error $\delta \theta_{\rm{conjugate}}$ as a function of total energy error, $\delta E$ and window size, $L$ for a fiducial conjugate direction of the H$_4$ square using a bound of $\delta E=0.001$. The black line is the isocurve of constant parameter error, and the black point corresponds to the optimal window size and energy error. \textbf{b}: Quadratic and quartic fits to the line search of an example direction for the H$_4$ square with target energy error $\delta E = 10^{-3}$ Hartree. The inset indicates the difference between the quadratic and quartic fits.
    \textbf{c}: Energy convergence of the line search algorithm vs. function calls for H$_4$ square with an error rate $\delta E = 10^{-4}$}
    \label{fig:h4squaredemo}
\end{figure}
\section{Simulations for Chemistry Models}
\label{sec:chemsimswalkthrough}

We will use the H$_4$ square as the example case to detail the entire optimization procedure.
The cases for N$_2$, H$_2$O, and the H$_4$ chain follow similarly.
The optimization procedure consists of five steps: finding a surrogate solution, calculating the Hessian, optimizing search windows for energy constraints, evaluating via a line search, and updating this line search.
We set the sparse wavefunction simulator (SWS) to keep 50 Slater determinants and optimize the system across all 193 parameters.
At this point, the Hessian is calculated numerically and then diagonalized.
We found that one of the 193 eigenvalues was within $10^{-4}$ of zero and eliminated this direction from the search.

The next step used is an optimization procedure to minimize the total energy error along an isocurve of constant parameter error, \(\delta \theta_{\text{conjugate}}\), as shown in panel a of Fig. \ref{fig:h4squaredemo}. The window optimization is carried out across all conjugate directions used. Once the search window sizes are optimized, we perform noisy line searches along each conjugate direction.
A quartic polynomial is then fit to the sampled energies along the line search directions. An example of these fits is shown in Fig. \ref{fig:h4squaredemo}.b.
The minimum for each conjugate direction is then used as the starting point for the next round of line searches. This is then repeated until satisfactory convergence is reached. We show the convergence for an error rate $\delta E = 10^{-4},10^{-5}$ in panel c of Fig. \ref{fig:h4squaredemo}. $N_{\rm{calls}}$ denotes the number of calls to the quantum computer normalized by the number of shots. Additionally, $E_{\rm{exact}}$ is the expected minimum of the VQE ansatz.

\section{Shot Count Analysis}

There are many proposed algorithms for reducing the number of shots required in energy evaluation \cite{ibm17tpbs,qwc20,chong20fullcommute,huang2020predicting,Crawford21si,dalfavero2023kqwc}. These methods have trade-offs between additional circuit depth, total shot count, classical processing complexity, and implementation difficulty. In this work we estimated the number of required shot counts using an approach based on the SortedInsertion algorithm of Crawford et al. \cite{Crawford21si}, a strategy that provides a reasonable balance between these trade-offs. 

First, during the surrogate coupled cluster calculations, we calculated the exact number of shots required for the surrogate wavefunction under the assumption that every term in the Hamiltonian $H = \sum a_i P_i$ is calculated separately on the quantum computer. The resulting quantity is
\begin{equation}
N_{\text{ungrouped}} = \frac{1}{\epsilon^2} \left ( \sum_i^M a_i \sqrt{\text{Var}[P_i]} \right )
\end{equation}
where $\text{Var}[P_i] = 1 - \langle P_i \rangle$, because each $P_i$ is a Pauli string. But in practice, one can simultaneously measure multiple commuting terms at the same time. Hence, $N_{\text{ungrouped}}$ is a worst-case ``baseline'' for the number of shots. This baseline calculation allows one to obtain shot count estimates for different term grouping strategies, even after the surrogate simulation.

To obtain a partitioning of the Pauli terms into commuting sets, we consider full commutation (as opposed to qubit-wise commutation \cite{qwc20}) before sorting the Pauli terms into sets using the \texttt{SortedInsertion} algorithm. This produces a partitioning into $O_k$ operators each with $m_k$ terms, where each $O_k = \sum_l^{m_k} a_{kl} P_{kl}$ and the $a_i$ have simply been relabeled as $a_{kl}$. We then calculate the scalar quantity $\hat R$ \cite{Crawford21si},

\begin{equation} \label{eqn:rhat}
    \hat{R} := \left[ 
     \frac{\sum_{k = 1} \sum_{l = 1}^{m_k} \left| a_{kl} \right|}
     {\sum_{k = 1} \sqrt{ \sum_{l = 1}^{m_k} \left| a_{kl} \right| ^ 2}}
    \right] ^ 2 .
\end{equation}
$\hat R$ has been shown to provide a reasonable estimate for reducing shot counts, with respect to the worst-case $N_{\text{ungrouped}}$. Hence, our final estimate in total shot counts is reported as $N_{\text{shots}} = N_{\text{ungrouped}}/ \hat{R}$. For our two smaller Hamiltonians (H$_2$O/STO-3G and N$_2$/STO-3G), this partitioning strategy led to approximately an order of magnitude reduction in shot counts. In comparison, for the two larger Hamiltonians (N$_2$/cc-pVDZ and H$_4$/cc-pVDZ) the decrease was approximately two orders of magnitude.




\section{Quantum Simulations}
\subsection{Quantum noise and error mitigation}
\label{sec:quantumnoise}

Real-life implementation of quantum algorithms will inevitably suffer from the effects of noise.
The ideal model of usual quantum computation involves sampling from probability distributions that can be modeled as originating (via Born's rule \cite{Peres2002}) from pure input states that are evolved through unitary dynamics, and followed by projective measurements \cite{Peres2002}.
However, in the presence of noise, the distribution that is sampled on a quantum device (for a fixed circuit) can significantly differ from the ideal, and in general should be described using more general formalism of mixed states, non-unitary dynamics (quantum channels \cite{nielsen2010quantum} for gate-based model and Lindbladians for continuous-time description \cite{nielsen2010quantum}), and generalized measurements \cite{Peres2002}.

In the context of variational optimization, one usually uses measurement results (samples) to construct a cost function that is an estimator of some quantity of interest, such as the expected value of the energy of the Hamiltonian evaluated on the variational state.
Modeling the effects of noise on the energy landscapes for variational optimization is an active area of research \cite{Barron2021,Xue2021,wang2021noise,Marshall2020}, and not many analytical results are currently known, except for the simplest noise models, such as global depolarizing noise that simply flattens the landscape (by rescaling the energy values) \cite{Urbanek2021}. 

The field of error mitigation (EM) tackles the problem of reducing the effects of noise on the results of quantum computation.
While some error-mitigation methods aim to improve the \textit{samples} from the probability distribution of interest (so-called strong error-mitigation), a simpler task consists of constructing error-mitigated \textit{estimators} of the quantity of interest, such as energy (so-called weak error-mitigation).
The main goal of weak error mitigation is usually to arrive at estimators for which the mean is closer to the ideal value, often at the cost of increased variance (see, for example, \cite{quek2023exponentially} for more discussion).

Let us now briefly discuss the EM techniques used in the main text (see Figs. \ref{fig:ibmbrisbaneresults} and \ref{fig:noisefloor}), which include Randomized Compilation (RC) \cite{PhysRevA.94.052325,Erhard_2019,li2017efficient, 2018efficienttwirling, 2013PhRvA..88a2314G,Silva-PT2008,Winick:2022scr}, Dynamical Decoupling (DD) \cite{PhysRevLett.82.2417, Ezzell2022,Carr1954,Meiboom1958,Maudsley1986,Viola2003}, Readout Error Mitigation (REM) \cite{Maciejewski2020,Maciejewski2021,van_den_Berg_2022}, Zero Noise Extrapolation (ZNE) \cite{PhysRevLett.119.180509,Kandala_2019, PhysRevX.7.021050,GiurgicaTiron2020}, Probabilistic Error Cancellation (PEC) \cite{Temme:2016vkz} , as well as Clifford Rescaling \cite{Urbanek2021,Vovrosh:2021ocf,Arahman:2022tkr}.

\textbf{Randomized compilation} (RC) is a strong EM technique that aims to reduce the effects of coherent noise on the circuit.
The technique assumes that available gates can be divided into "easy" (usually 1-qubit gates) and "hard" (usually 2-qubit gates) sets.
A subset of easy gates (twirling subset) is then inserted randomly in the circuit layers (layers are defined w.r.t. the implementation of hard gates).
The insertion of gates is designed in a way that the effective noise channel acting on hard gates is transformed into Pauli noise, thus removing coherent noise effects.
Typically, the twirling subset consists of single-qubit Pauli gates -- indeed, this is the subset used in our experiments.
Note that RC requires circuit randomization which does not generate sample overhead, but might be hard to implement in practice.

\textbf{Dynamical Decoupling} (DD) is a strong EM method that aims to reduce the noise processes (usually dephasing) affecting idle qubits, i.e., qubits on which, in a given moment, no quantum gates act.
It consists of applying periodic sequences of single-qubit pulses to the idle qubits to effectively average out spurious interactions, thus reducing decoherence and, in some variants, cross-talk (typically ZZ interactions). 
There are multiple variants of DD \cite{PhysRevLett.82.2417, Ezzell2022,Carr1954,Meiboom1958,Maudsley1986,Viola2003}, in this work we use the $XY-4$ sequence.
Note that DD does not require the implementation of additional circuits, and does not generate additional sample overhead.

\textbf{Readout Error Mitigation} (REM) is a general term to describe techniques that reduce the effects of the measurement noise on the estimated quantities.
Although a more general treatment might sometimes be required (see, e.g., \cite{tuziemski2023efficient}), the measurement noise is usually modeled as a stochastic map $\Lambda$ applied to the probability distribution $\mathbf{p}^{\left(\mathrm{input}\right)}$ that would have been obtained if the quantum measurement was perfect, i.e., we obtain noisy distribution $\mathbf{p}^{\left(\mathrm{noisy}\right)} = \Lambda\mathbf{p}^{\left(\mathrm{input}\right)}$.
Since $\mathbf{p}^{\left(\mathrm{noisy}\right)}$ can be estimated directly from measurement samples obtained from a device, the simplest REM method consists of applying $\Lambda^{-1}$ to the estimated distribution (and projecting onto the space of probability distributions in case of unphysical entries, see, e.g., \cite{Maciejewski2020}).
Some methods exist that tackle scalability issues and are based on applying $\Lambda^{-1}$ on a reduced subspace that is in some proximity to the observed measurement results (this requires the assumption that measurement noise is sufficiently weak) \cite{Nation2021}.
In this work, we use Qiskit Runtime's implementation of the Twirled Readout Error eXtinction (TREX) method \cite{van_den_Berg_2022}.
This REM technique symmetrizes the effective noise model $\Lambda$ by introducing random combinations of $X$ gates followed by appropriate post-processing.
Due to twirling, the noise effectively rescales the expected values of observables.
This rescaling can be characterized and reversed in the post-processing, at the cost of increased variance.

\textbf{Zero Noise Extrapolation} (ZNE) is a weak error-mitigation strategy that consists of estimating a given quantity (such as energy) for multiple effective noise strengths (typically quantified by a single parameter $\gamma^{\left(\mathrm{ZNE}\right)}$), and extrapolating the results to the zero-noise regime.
To perform ZNE, one needs 1) A way to increase noise strength in a controllable manner, and 2) A choice of fitting model for extrapolation. 
One of the standard methods to perform 1) is so-called unitary-folding, where multiple sequences of the circuit and its inverse are appended to the circuit. 
Ideally, they should not change anything, but in the presence of noise, the additional gates will increase the noise strength \cite{GiurgicaTiron2020}.
In \cite{PhysRevX.7.021050} a method to increase a certain type of local Pauli noise (so-called sparse Pauli–Lindblad model, see \cite{vandenBerg2023}) was introduced that uses random insertions of Pauli gates.
The 2) are usually chosen to be exponential decay or low-degree polynomials.
The simplest example is linear regression $E\left(\gamma^{\left(\mathrm{ZNE}\right)}\right) = a\gamma^{\left(\mathrm{ZNE}\right)} +b$, where it is easily seen that the zero-noise value can be inferred as equal to the intercept $b$.
In this work, we use a quadratic ansatz, $E(\gamma^{\rm{(ZNE)}}) = a (\gamma^{\rm{(ZNE)}})^2 + b \gamma^{\rm{(ZNE)}} + c$
Note that ZNE requires implementing multiple variants of the same circuit with changing noise strengths, thus introducing a constant multiplicative overhead in the number of circuits (in the special case of ZNE tailored to the Pauli-Lindblad model, it also requires randomized circuits' implementation).
Furthermore, the constructed estimator can have a bias and an increased variance which generally introduces additional sample overheads.

\textbf{Probabilistic Error Cancellation} (PEC) is a weak EM method that decomposes a (possibly approximate) inverse of the noise channel (which, in general, is not a physical map) into a linear combination of some set of implementable physical channels. 
The coefficients in the decomposition are given by quasiprobability distribution $\left\{q_i\right\}$ (with $|q_i|\leq 1$, $\sum_{i}q_i=1$) which is used to sample the available physical channels and to construct estimators of quantities of interest from gathered statistics.
To perform PEC, one needs 1) Choose a model for the noise channel to invert, and 2) Find a quasiprobability decomposition of the channels' inverse.
To perform 1), a standard process tomography \cite{nielsen2010quantum} or gate-set tomography \cite{Greenbaum:2015aqm} can be used. 
In the case of Pauli noise, methods tailored to its reconstruction can be more efficient \cite{Flammia2020}.
For the special case of the sparse Pauli–Lindblad model mentioned previously, a method based on Pauli insertions was proposed in \cite{PhysRevX.7.021050}.
In this work, we use Qiskit's implementation of such PEC.
Note that PEC requires noise characterization and randomized circuit implementation.
The constructed estimators are (ideally) unbiased, but they have an increased variance that scales like $\gamma^{\left(\mathrm{PEC}\right)} = \sum_{i}|q_i|$, which introduces sample complexity overheads. 
Since PEC is usually performed independently for local gates, the total overhead scales exponentially in circuit size.

\textbf{Clifford rescaling} is perhaps one of the simplest weak EM methods available. 
The technique approximates the noise model as a global depolarizing channel acting just before measurement, characterized by error probability $p^{\left(dep\right)}$.
The effect of the global depolarizing channel on the expected value of energy is a simple linear transformation that for traceless observables (such as Pauli Hamiltonians) reduces to rescaling. 
The error-mitigated estimator is then constructed via multiplication by $\frac{1}{1-p^{\left(dep\right)}}$.
Note that rescaling an estimator (by a number greater than 1) increases variance.

\begin{figure}[!ht]
\includegraphics[width=\linewidth]{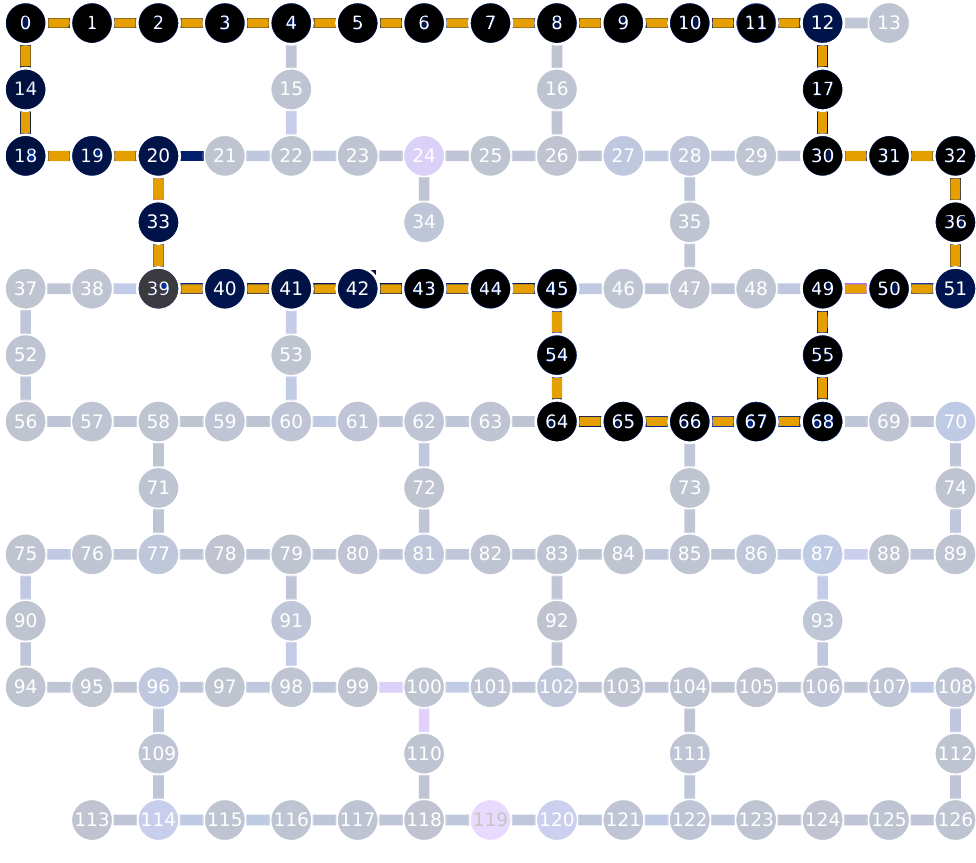}
\caption{Topology for the $\texttt{ibm\_brisbane}$ quantum processing unit. The orange markers denote the path of qubits used for the line search calculations of the transverse Ising model. The faded out qubits and connections indicate unused qubits and connections. The graphic is used from \cite{ibmwebpage}}
\label{fig:dabrisbanemachine}
\end{figure}
\subsection{Line Search Calculations}
The line search calculations 
performed on the $\texttt{ibm\_brisbane}$ quantum processing unit for the transverse Ising model Hamiltonian are discussed in this section. The topology for the quantum hardware is provided in Fig. \ref{fig:dabrisbanemachine}. Additional hardware specifics are provided in an a supplemental json file. 
The model Hamiltonian we use is 
\begin{equation}
    \label{eq:hamsup}\hat{H} = J_1 \sum_{i=1}^{N_s}\hat{\sigma}^z_i\hat{\sigma}^z_{i + 1} + J_2 \sum_{i=1}^{N_s}\hat{\sigma}^z_i \hat{\sigma}^z_{i + 2} + h_t \sum_{i=1}^{N_z} \hat{\sigma}^x_i.
\end{equation}
where the parameters $J_1 = 1.0$, $J_2 = 0.9$, $h_t = 0.4$, and the number of sites, $N_s = 40$. We use periodic boundary conditions. 
We use a unitary coupled cluster like ansatz shown in Fig. \ref{fig:ansatz}.
The ansatz in explicit equation form is
\begin{figure}[!t]
\begin{gather*}
    \Qcircuit @R=1em @C=1em {
    & \vdots & & \multigate{1}{e^{-i\theta_2(\hat{\sigma}^y_{N_s - 1}\hat{\sigma}^z_0 - \hat{\sigma}^z_{N_s - 1}\hat{\sigma}^y_0)}} & \qw & \vdots & & \multigate{1}{e^{-i\theta_4(\hat{\sigma}^y_{N_s - 1}\hat{\sigma}^z_0 - \hat{\sigma}^z_{N_s - 1}\hat{\sigma}^y_0)}} & \qw & \vdots\\
    & \multigate{1}{e^{-i\theta_1 (\hat{\sigma}^y_0\hat{\sigma}^z_1 - \hat{\sigma}^z_0\hat{\sigma}^y_1)}} & \qw & \ghost{e^{-i\theta_2(\hat{\sigma}^y_{N_s - 1}\hat{\sigma}^z_0 - \hat{\sigma}^z_{N_s - 1}\hat{\sigma}^y_0)}} & \qw & \multigate{1}{e^{-i\theta_3 (\hat{\sigma}^y_0\hat{\sigma}^z_1 - \hat{\sigma}^z_0\hat{\sigma}^y_1)}} & \qw & \ghost{e^{-i\theta_2(\hat{\sigma}^y_{N_s - 1}\hat{\sigma}^z_0 - \hat{\sigma}^z_{N_s - 1}\hat{\sigma}^y_0)}} & \qw & \qw \\
    & \ghost{e^{-i\theta_1 (\hat{\sigma}^y_0\hat{\sigma}^z_1 - \hat{\sigma}^z_0\hat{\sigma}^y_1)}} & \qw & \multigate{1}{e^{-i\theta_2 (\hat{\sigma}^y_1\hat{\sigma}^z_2 - \hat{\sigma}^z_1\hat{\sigma}^y_2)}} & \qw & \ghost{e^{-i\theta_3 (\hat{\sigma}^y_0\hat{\sigma}^z_1 - \hat{\sigma}^z_0\hat{\sigma}^y_1)}} & \qw & \multigate{1}{e^{-i\theta_4 (\hat{\sigma}^y_1\hat{\sigma}^z_2 - \hat{\sigma}^z_1\hat{\sigma}^y_2)}} & \qw & \qw \\
    & \multigate{1}{e^{-i\theta_1 (\hat{\sigma}^y_2\hat{\sigma}^z_3 - \hat{\sigma}^z_2\hat{\sigma}^y_3)}} & \qw & \ghost{e^{-i\theta_2 (\hat{\sigma}^y_1\hat{\sigma}^z_2 - \hat{\sigma}^z_1\hat{\sigma}^y_2)}} & \qw & \multigate{1}{e^{-i\theta_3 (\hat{\sigma}^y_2\hat{\sigma}^z_3 - \hat{\sigma}^z_2\hat{\sigma}^y_3)}} & \qw & \ghost{e^{-i\theta_4 (\hat{\sigma}^y_1\hat{\sigma}^z_2 - \hat{\sigma}^z_1\hat{\sigma}^y_2)}} & \qw & \qw \\
    & \ghost{e^{-i\theta_1 (\hat{\sigma}^y_2\hat{\sigma}^z_3 - \hat{\sigma}^z_2\hat{\sigma}^y_3)}} & \qw & \multigate{1}{e^{-i\theta_2 (\hat{\sigma}^y_1\hat{\sigma}^z_2 - \hat{\sigma}^z_1\hat{\sigma}^y_2)}} & \qw & \ghost{e^{-i\theta_3 (\hat{\sigma}^y_2\hat{\sigma}^z_3 - \hat{\sigma}^z_2\hat{\sigma}^y_3)}} & \qw & \multigate{1}{e^{-i\theta_4 (\hat{\sigma}^y_1\hat{\sigma}^z_2 - \hat{\sigma}^z_1\hat{\sigma}^y_2)}}  & \qw & \qw \\
    & \vdots &  & \ghost{e^{-i\theta_2 (\hat{\sigma}^y_3\hat{\sigma}^z_4 - \hat{\sigma}^z_3\hat{\sigma}^y_4)}} & \qw & \vdots & &\ghost{e^{-i\theta_2 (\hat{\sigma}^y_3\hat{\sigma}^z_4 - \hat{\sigma}^z_3\hat{\sigma}^y_4)}} & \qw & \vdots \\
    }
\end{gather*}
\caption{Circuit depiction of the ansatz for Eq. (\ref{eq:ansatzsupp})}
\label{fig:ansatz}
\end{figure}
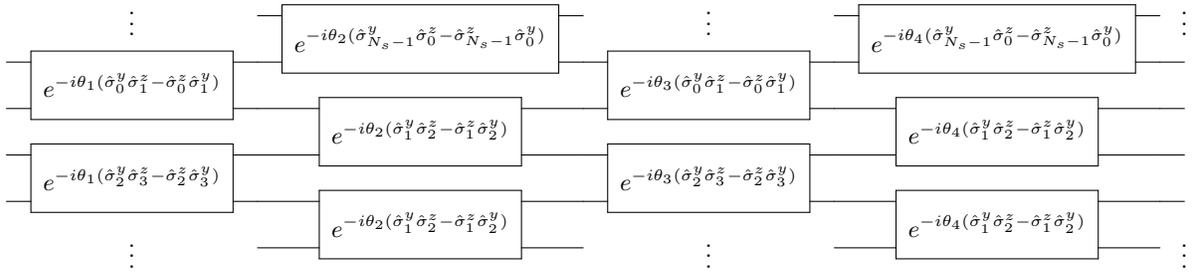
\begin{equation}
\label{eq:ansatzsupp}
    \hat{V}(\vec{\theta}) = \Bigg(\prod_{i=0}^{N_s / 2}\hat{U}_{2i, 2i + 1}(\theta_1)\Bigg)\Bigg(\prod_{i=0}^{N_s / 2} \hat{U}_{2i + 1, 2i + 2}(\theta_2)\Bigg)
    \Bigg(\prod_{i=0}^{N_s / 2}U_{2i, 2i + 1}(\theta_3)\Bigg)\Bigg(\prod_{i=0}^{N_s / 2} \hat{U}_{2i + 1, 2i + 2}(\theta_4)\Bigg),
\end{equation}
where
\begin{equation}
    \hat{U}_{i, j}(\theta) = e^{-i \theta (\hat{\sigma}^y_i\hat{\sigma}^z_j - \hat{\sigma}^y_j\hat{\sigma}^z_i)}.
\end{equation}
Since the model in principle should be translationally invariant we force all identical operators implemented in parallel to have the same angle.
For a gradient based optimization this may lead to a high sampling overhead for calculations of derivatives. However for a gradient free method, such as the line search, this is less problematic.

First we look for the best parameters from the surrogate model.
In this case a matrix product state simulator was used to simulate the ansatz given in Eq. (\ref{eq:ansatzsupp}). 
This simulator used a bond dimension of 4.
While the ansatz itself is simple enough to calculate with a classically tractable bond dimension of 100, this artificially small bond dimension was used to provide a proof of principle demonstration on a quantum processor.

\begin{figure}[!t]
    \centering
    \includegraphics[width=\linewidth]{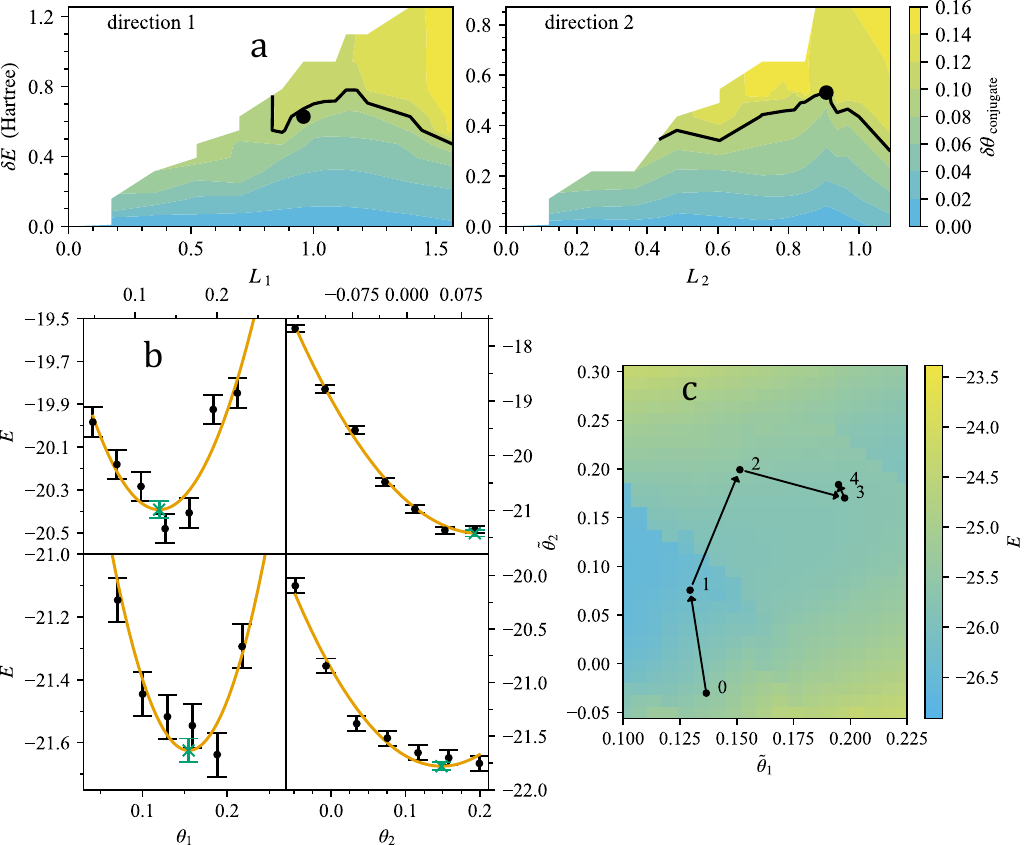}
    \caption{Summary of the key aspects of the surrogate line search optimization for the transverse Ising model from Eq. (\ref{eq:hamsup}). Panel \textbf{a} shows the search window optimization for conjugate parameters $\tilde{\theta}_1$ and $\tilde{\theta}_2$. 
    Panel \textbf{b} shows the first (top panes) and second (bottom panes) iterations of the line search. The orange line is the quadratic fit used to calculate the minimum along the search direction. The black points correspond to the energies calculated on $\texttt{ibm\_brisbane}$, the orange curve corresponds to the quadratic fit, and the green cross corresponds to the ansatz minimum. Panel \textbf{c} shows the movement of the conjugate parameters (denoted by black circles) through the potential energy surface for Eq. \ref{eq:ansatzsupp}.}
    \label{fig:qpudemo}
\end{figure}

After finding the optimal solution, the Hessian in parameter space was calculated. The Hessian has eigenvalues: 246.45, 118.49, 8.58, -8.26.
Given the large spread between the two largest eigenvalues and the two smallest one, along with a negative eigenvalue we decided to truncate to a line search along 2 conjugate directions. These directions are denoted, $\tilde{\theta}_1$ and $\tilde{\theta}_2$. The search windows for $\tilde{\theta}_1$ and $\tilde{\theta}_2$ were optimized for an energy uncertainty of $\delta E = 0.01$. The optimization is shown in Fig. \ref{fig:qpudemo}a.

The step of performing the line searches was done asynchronously to minimize latent time on the quantum processing unit. The line searches along with all necessary error mitigation circuits were submitted as a batch job to $\texttt{ibm\_brisbane}$. When all the calculations in the job were received, the fits to the various line searches were performed. A subset of these corresponding to the first and second iterations are shown in Fig. \ref{fig:qpudemo}b. 
For all of the quantum processing unit simulations we used the following error mitigation techniques: dynamical decoupling \cite{PhysRevLett.82.2417, Ezzell2022,Carr1954,Meiboom1958,Maudsley1986,Viola2003}, randomized compiling \cite{PhysRevA.94.052325,Erhard_2019,li2017efficient, 2018efficienttwirling, 2013PhRvA..88a2314G,Silva-PT2008,Winick:2022scr}, and Clifford rescaling\cite{Urbanek2021,Vovrosh:2021ocf,Arahman:2022tkr}.
A brief overview of these and other error mitigation methods can be found in Sec. \ref{sec:quantumnoise}.
The circuits used the XY-4 decoupling sequence innately provided by the \texttt{qiskit\_runtime} library. 

The movement of the conjugate parameters through the potential energy surface are shown in Fig. \ref{fig:qpudemo}c. While in principle the optimization should move relatively smoothly toward the global minimum. 
However the quantum noise can introduce certain distortions to the potential energy surface including biases from coherent unitary errors. Nevertheless the optimization appears to converge to a reasonable result given the possible distortion to the potential energy landscape. 
This converged result is within 10$\%$ of the expected minimum from the ansatz. 

\subsection{Energy calculation on a QPU}
\label{app:energy_calculation_on_QPU}
We calculated the energy associated with the variational circuits using the parameters corresponding to the surrogate minimum of the transverse Ising model using the $\texttt{ibm\_brisbane}$ QPU. The data were obtained between 12-10-2023 and 01-01-2024. We error-mitigated the results shown in Fig. \ref{fig:noisefloor} in the main text with dynamic decoupling, readout error mitigation, digital zero-noise extrapolation, and probabilistic error cancellation using Qiskit IBM Runtime implementation \cite{Qiskit} all of which are outlined in the prior section. The data points in the figure are averaged over four identical QPU runs, each targeted for 8192 shots. (The exact number of shots is automatically adjusted by the error mitigation algorithms such that the statistical error after error mitigation is estimated to be consistent with the target of 8192 shots.) The error bars are given by one standard deviation of the four QPU runs.
The qubits used for each system size are listed in Tab. \ref{tab:qubitlayouts}. Calibration data for the various simulations are provided in an attached JSON file. 

\begin{table}[!ht]
    \centering
    \begin{tabular}{cc}\hline\hline
         $N_s$ & Qubits  \\\hline
         12 & 0-4, 14, 15, 18-22\\ 
         20 & 0-8, 16, 18-26\\
         28 & 0-12, 14, 17-30\\
         36 &  8-12, 16, 17, 26-28, 30-32, 35, 36, 41-47, 49-51, 53, 55, 60-68\\
         44 &  0-12, 14, 17-28, 30-32, 35, 36, 45-47, 49-51, 54, 55, 64-68\\
         52 & 0-8, 53, 41-43, 34, 18-24, 14, 16, 26-32, 36, 49-51, 55, 58-60, 66-68, 71, 73, 77-85 \\
         \multirow{2}*{60} & 4-12, 14, 15, 17, 22-24, 28-30, 34, 35, 37-43, 47-49, 52, 55-58, 66-68, 71, 73, 75-77, 85-83, 90, 92, 94-96, 100-102, 109,\\
         & 110, 114-118\\
         \multirow{2}*{68} & 0-12, 17-24, 28-30, 34, 35, 37-43, 45-47, 52, 54, 56-58, 62-64, 71, 72, 75-77, 79-81, 90, 91, 94-96, 98-100, 109, 110,\\
         & 114-118\\
         76 & 0-12, 14, 17-24, 28-30, 34, 35, 37-43, 47-49, 52, 55-62, 66-68, 72, 73, 75-81, 83-85, 90, 92, 94-96, 100-102, 109, 110, 114-118\\
         84 & 0-12, 14, 17-20, 30-33, 36-39, 45-52, 54, 56-58, 64-66, 71, 73, 75-77, 79-84, 85, 90, 91, 94-96, 98-109, 112, 114-126\\\hline
    \end{tabular}
    \caption{Qubit usage for various noise studies. Qubits are listed in numerical order.}
    \label{tab:qubitlayouts}
\end{table}

The QPU results are compared to the MPS result with a bond dimension 40. While the circuit depth remains constant with increasing system size $N_{s}$, the number of two-qubit and measurement gates increases linearly with $N_{s}$. The two-qubit gate and readout errors account for the linear treads of the deviation between the QPU and MPS results. These two error sources are partially mitigated by readout-error mitigation and PEC, with noticeable improvements shown in the figure. The qubit decoherence is not a limiting factor in this calculation, as suggested by the fact that ZNE does not improve the results.

\bibliographystyle{apsrev4-1}

%

\end{document}